# Measurement of Magnonic Quantum Geometry

S. Kim[1], M. Stone[2], S. E. Nagler[2,3], A. D. Christianson[1], Jiabin Yu[4,*], S. Okamoto[1,*], H. Miao[1,*]

[1]Materials Science and Technology Division, Oak Ridge National Laboratory, Oak Ridge, Tennessee 37831, USA

[2]Neutron Scattering Division, Oak Ridge National Laboratory, Oak Ridge, Tennessee 37831, USA

[3]Department of Physics and Astronomy, University of Tennessee, Knoxville, TN 37996, USA

[4]Department of Physics, University of Florida, Gainesville, Florida 32611, USA

**Quantum geometry plays a fundamental role in macroscopic quantum phenomena, including quantum Hall effects and flat-band superconductivity. However, since quantum geometry is encoded in wavefunctions rather than energy spectra, direct experimental characterization remains a formidable challenge. This challenge is particularly acute for quasiparticles in quantum materials, whose wavefunctions are generally inferred from effective low-energy models rather than measured directly. Here, we establish an experimental framework for characterizing the quantum geometry of magnons, a ubiquitous Bosonic quasiparticle of magnetically ordered states. We show that the inelastic neutron scattering (INS) cross-section measures geometric parameters of the spinor wavefunctions of ferromagnets. By tracking the energy- and momentum-dependent INS intensity, we extract the quantum geometric tensor without requiring detailed knowledge of microscopic spin-spin interactions. Our results establish magnons as an ideal arena for experimentally probing quantum geometry and open new avenues for exploring geometry-driven magnon transport and quantum Fisher information in magnetic materials.**

## Introduction:

Geometry underpins our understanding of the physical world, from Einstein's description of gravity to emergent phases of quantum materials. In solid states, the geometry of Bloch wavefunctions in momentum space is encoded in the quantum geometric tensor (QGT), whose antisymmetric and symmetric parts correspond to the Berry curvature, $\boldsymbol{\Omega}$, and quantum metric, $\boldsymbol{g}$, respectively.[1-6] The Berry curvature,[7,8] that acts as an effective magnetic field, underlies a broad range of topological phenomena, including quantum Hall effects[9,10] and topological superconductivity.[11-13] The quantum metric, which quantifies the distance between neighboring quantum states, has recently emerged as a fundamental quantity governing flat-band superconductivity,[14-16] nonlinear transport responses,[17-21] orbital magnetic susceptibility,[22-24] and chiral magnets.[25,26]

Despite its fundamental importance, direct experimental access to the full QGT has remained extremely limited. To date, complete measurements have only been realized in engineered two-level quantum systems, such as superconducting qubits[27], nitrogen-vacancy centers[28], and microcavity.[29] Extraction of the electronic QGT from quantum materials was achieved very recently using angle-resolved photoemission spectroscopy in bulk black phosphorus using a two-band approximation and space-time inversion symmetry.[30] However, since electronic Bloch states generally involve orbital degree of freedom and extend over a large energy scale,[31] the experimentally reconstructed QGT is approximate and captures only a projected subset of the Bloch wavefunction, $|\boldsymbol{\psi}_n(\boldsymbol{k})\rangle$, where $n$ and $\boldsymbol{k}$ are band index and crystal momentum, respectively.

Here we show that magnons provide an ideal Bosonic platform for the measurement of QGT. In a ferromagnetic (FM) system with two spins in a unit cell, the sublattice degree of freedom gives rise to an emergent pseudospin-1/2 and realizes an intrinsic two-level quantum system in THz-energy scale while remaining embedded in a crystalline material. We demonstrate that the momentum-dependent pseudospin texture enters directly into the cross-section of inelastic neutron scattering (INS), enabling the experimental determination of the complete magnonic QGT. Applying this approach to the elemental ferromagnet Gd, we construct the magnonic QGT and reveal pronounced nontrivial quantum metric around the K point and trivial QGT near the Γ point. Our work introduces collective spin excitations as a powerful new platform for exploring quantum-geometric phenomena in condensed matter systems.

**Quantum geometry of magnons:**

We first introduce the magnonic QGT. The generic single-band Fubini-Study QGT is defined as:

$$Q_n^{ij}(\boldsymbol{k}) = \langle \partial_{k_i} u_n(\boldsymbol{k}) | 1 - (|u_n(\boldsymbol{k})\rangle\langle u_n(\boldsymbol{k})|) | \partial_{k_i} u_n(\boldsymbol{k}) \rangle, \quad (1)$$

where $|u_n(\boldsymbol{k})\rangle = e^{-\mathrm{i}\boldsymbol{k}\cdot\boldsymbol{r}}|\psi_n(\boldsymbol{k})\rangle$ is the periodic part of the $n^{\text{th}}$ Bloch wavefunction. The indexes $i$ and $j$ run over the coordinate of the momentum space. For FM systems with $m$ local spins in a unit cell, a one-magnon Bloch state at each $\boldsymbol{k}$ point has $m$ components (see Supplementary Materials S1). As an example, Fig. 1**a** depicts a simple 1D FM spin chain with one local spin in the unit cell, i.e., $m$=1. This leads to zero magnon QGT and hence trivial quantum geometry. Non-trivial QGT naturally emerge in magnetic systems with more than one local spin in the unit cell. Figure 1**b** illustrates collective spin precessions in a FM honeycomb lattice. Like the electronic structure of graphene,[9] the magnetic A-sublattice (cyan) and B-sublattice (orange) yield effective pseudospin-1/2 (see Supplementary Materials S2) and give rise to the generic magnon Hamiltonian:[32]

$$H(\boldsymbol{k}) = d_0(\boldsymbol{k})\, I + \boldsymbol{d}(\boldsymbol{k}) \cdot \boldsymbol{\sigma}, \quad (2)$$

where $I$ is the 2×2 identity matrix and $\boldsymbol{\sigma} = (\sigma_x, \sigma_y, \sigma_z)$ are Pauli matrices. The vector $\boldsymbol{d}(\boldsymbol{k})$ acts as a pseudo-magnetic field and can be parametrized as $\boldsymbol{d}(\boldsymbol{k}) = |\boldsymbol{d}(\boldsymbol{k})|\hat{\boldsymbol{d}}(\boldsymbol{k})$ with $\hat{\boldsymbol{d}}(\boldsymbol{k}) = (\sin(\theta_{\boldsymbol{k}})\cos(\varphi_{\boldsymbol{k}}), \sin(\theta_{\boldsymbol{k}})\sin(\varphi_{\boldsymbol{k}}), \cos(\theta_{\boldsymbol{k}}))$. Here $\theta_{\boldsymbol{k}}$ and $\varphi_{\boldsymbol{k}}$ are geometric parameters on the Bloch sphere (Figs. 1**c** and **e**). Then, the corresponding two-component complex spinor wavefunctions are $u_+(\boldsymbol{k}) = \begin{pmatrix} \cos\left(\frac{\theta_{\boldsymbol{k}}}{2}\right) \\ e^{i\varphi_{\boldsymbol{k}}}\sin\left(\frac{\theta_{\boldsymbol{k}}}{2}\right) \end{pmatrix}$ and $u_-(\boldsymbol{k}) = \begin{pmatrix} -\sin\left(\frac{\theta_{\boldsymbol{k}}}{2}\right) \\ e^{i\varphi_{\boldsymbol{k}}}\cos\left(\frac{\theta_{\boldsymbol{k}}}{2}\right) \end{pmatrix}$, for the magnon dispersion $\omega_{+,\boldsymbol{k}} = d_0(\boldsymbol{k}) + |\boldsymbol{d}(\boldsymbol{k})|$ and $\omega_{-,\boldsymbol{k}} = d_0(\boldsymbol{k}) - |\boldsymbol{d}(\boldsymbol{k})|$, respectively. At the Brillouin zone center, the – and + spinors are corresponding in-phase acoustic magnon and out-phase optical magnon, respectively. Following Eq. (1), the magnonic QGT can be expressed in terms of the $\hat{\boldsymbol{d}}(\boldsymbol{k})$-vector as:

$$g_{\pm}^{ij}(\boldsymbol{k}) = \frac{1}{4}\partial_{k_i}\hat{\boldsymbol{d}}(\boldsymbol{k}) \cdot \partial_{k_j}\hat{\boldsymbol{d}}(\boldsymbol{k}), \quad (3)$$

$$\Omega_{\pm}^{ij}(\boldsymbol{k}) = \mp\frac{1}{2}\hat{\boldsymbol{d}}(\boldsymbol{k}) \cdot \left[\partial_{k_i}\hat{\boldsymbol{d}}(\boldsymbol{k}) \times \partial_{k_j}\hat{\boldsymbol{d}}(\boldsymbol{k})\right], \quad (4)$$

**Measurement of magnonic quantum geometry**

Since the pseudospin-1/2 emerges from the sublattice degree of freedom, the spinor wavefunction $u_{\pm}(\boldsymbol{k})$ is expected to directly enter the cross-section of scattering probes. Indeed, for a two-band FM magnon system, the INS intensity can be written as (see Supplementary Materials S3):

$$I(\boldsymbol{Q}, \omega) = C(\boldsymbol{Q}) \sum_{n=\pm}[1 + n\sin(\theta_{\boldsymbol{k}})\cos(\varphi_{\boldsymbol{k}} + \varphi_{\boldsymbol{G}})]\, \delta(\omega - \omega_{n,\boldsymbol{k}}), \quad (5)$$

where $C(\boldsymbol{Q})$ is a positive factor that contains Lande factor, magnetic form factor and the polarization factor. The total momentum transfer, $\boldsymbol{Q} = \boldsymbol{p}_i - \boldsymbol{p}_f$ with $\boldsymbol{p}_i, \boldsymbol{p}_f$ the initial and final neutron momentum. The magnon eigenvalue, $\omega_{n,\boldsymbol{k}}$, is defined at Bloch momentum, $\boldsymbol{k} = \boldsymbol{Q} - \boldsymbol{G}$, where $\boldsymbol{G} = (H, K, L)$ in reciprocal lattice unit (r.l.u.) with integer $H$, $K$, and $L$. The phase $\varphi_{\boldsymbol{G}} = \boldsymbol{G} \cdot \boldsymbol{r}_{AB}$ arises from the scattering phase difference between A- and B-sublattices, where $\boldsymbol{r}_{AB} = \boldsymbol{r}_B -$ $\boldsymbol{r}_A$ is the real space translation vector from A-sublattice to B-sublattice (Fig. 1**b**). Integrating $\omega$ around $\omega_{n,\boldsymbol{k}}$ gives:

$$I_n(\boldsymbol{Q}) = \int_{\omega_{n,\boldsymbol{k}}+0^-}^{\omega_{n,\boldsymbol{k}}+0^+} d\omega\, I(\boldsymbol{Q},\omega) = C(\boldsymbol{Q})[1 + n\sin(\theta_{\boldsymbol{k}})\cos(\varphi_{\boldsymbol{k}} + \varphi_{\boldsymbol{G}})], \quad (6)$$

The total INS intensity at $\boldsymbol{Q}$ is $I_+(\boldsymbol{Q}) + I_-(\boldsymbol{Q}) = 2\,C(\boldsymbol{Q})$. We can then introduce energy integrated and energy-resolved INS geometric quantities:

$$J(\boldsymbol{Q}) = \frac{I_+(\boldsymbol{Q}) - I_-(\boldsymbol{Q})}{I_+(\boldsymbol{Q}) + I_-(\boldsymbol{Q})} = \sin(\theta_{\boldsymbol{k}})\cos(\varphi_{\boldsymbol{k}} + \varphi_{\boldsymbol{G}}) = \left[\hat{d}_{\boldsymbol{x}}(\boldsymbol{k})\cos(\varphi_{\boldsymbol{G}}) - \hat{d}_{\boldsymbol{y}}(\boldsymbol{k})\sin(\varphi_{\boldsymbol{G}})\right], \quad (7\text{-}1)$$

$$J_{n=\pm}(\boldsymbol{Q},\omega) = \frac{I(\boldsymbol{Q},\omega)}{I_+(\boldsymbol{Q}) + I_-(\boldsymbol{Q})} = [1 + n\sin(\theta_{\boldsymbol{k}})\cos(\varphi_{\boldsymbol{k}} + \varphi_{\boldsymbol{G}})]\,\delta(\omega - \omega_{n,\boldsymbol{k}}), \quad (7\text{-}2)$$

Significantly, $J(\boldsymbol{Q})$ is solely determined by geometric parameters of the wavefunction and independent on material properties. Since $J(\boldsymbol{k} + \boldsymbol{G})$ at a fix $\boldsymbol{k}$-point is modulated by $\varphi_{\boldsymbol{G}}$, the $\hat{d}_{\boldsymbol{x}}(\boldsymbol{k})$ and $\hat{d}_{\boldsymbol{y}}(\boldsymbol{k})$ can be fully determined by measuring $J(\boldsymbol{k} + \boldsymbol{G}_1)$ and $J(\boldsymbol{k} + \boldsymbol{G}_2)$ as long as $\sin(\varphi_{\boldsymbol{G}_1} - \varphi_{\boldsymbol{G}_2}) \neq 0$ (see Method). The $\hat{d}_{\boldsymbol{z}}(\boldsymbol{k})$ can then be determined up to a global sign. Since $g_{\pm}^{ij}(\boldsymbol{k})$ is independent on the global sign of $\hat{d}_{\boldsymbol{z}}(\boldsymbol{k})$, the $g_{\pm}^{ij}(\boldsymbol{k})$ can thus be uniquely determined. The sign of $\Omega_{\pm}^{ij}(\boldsymbol{k})$ can be determined at high-symmetry points based on symmetries (see Method). Following this framework, the QGT can be entirely constructed from experimental measurement without prior knowledge of microscopic magnetic interactions. These results thus provide a generic and model independent measurement of magnonic quantum geometry.

**Magnonic geometric tensor of elemental Gd**

We apply the above formulism to the FM Gd ($T_C = 293$ K). As we will show below, Gd provides a particularly transparent platform for the measurement of magnonic quantum geometry, owing to its simple magnon spectrum and strong, well-resolved INS signal enabled by its large magnetic moment. The data presented in the manuscript are from recent INS studies of Dirac magnons in Gd.[33,34] The elemental Gd crystallizes in the hexagonal close-packed (*hcp*) structure with space group $P6_3/mmc$. The primitive unit cell contains two nearly isotropic Gd atoms occupying the Wyckoff 2c positions, (1/3, 2/3, 1/4) and (2/3, 1/3, 3/4) with $\boldsymbol{r}_{AB} = (\frac{1}{3}, -\frac{1}{3}, \frac{1}{2})$. Previous spin wave modeling of the magnon spectrum has revealed long-range magnetic interactions up to the 26$^{\text{th}}$ nearest neighbors.[34,35] Despite the complexity of the spin Hamiltonian, the two magnetic sublattices of Gd allow its magnon Hamiltonian to be described by Eq. (2). Importantly, since Gd has negligible spin-orbit coupling, the ferromagnetic Gd has spinless time-reversal and inversion symmetry, which make $\hat{d}_z(\boldsymbol{k})$ vanish in the whole Brillouin zone (see Supplementary materials S4). Therefore, the Berry curvature is zero throughout the Brillouin zone, and we only need to measure $\hat{d}_x(\boldsymbol{k})$ and $\hat{d}_y(\boldsymbol{k})$.

Figure 2**a** shows $J_{n=\pm}(\boldsymbol{Q}, \omega)$ along the (H, H, 0) direction. The $u_+(\boldsymbol{k})$ (blue) and $u_-(\boldsymbol{k})$ (red) modes can only be observed at even (red) and odd (blue) integer *L*, respectively. This *L*-dependent "dichroism" comes from the extra spinless glide symmetry along (H, H, 0), which further enforce $\hat{d}_y(\boldsymbol{k}) = 0$ and thus $\hat{d}_x(\boldsymbol{k}) = \pm 1$. As $\varphi_{\boldsymbol{G}} = L\pi$, the $\hat{d}_x(\boldsymbol{k}) = 1$ is picked by even $L$, while the $\hat{d}_x(\boldsymbol{k}) = -1$ is picked by odd L. The measurement clearly determines the $\hat{\boldsymbol{d}}(\boldsymbol{k})$ value on this high symmetry line.

To more clearly demonstrate the power of our method, Fig. 2**b** and **c** show experimentally determined $\hat{d}_x(\boldsymbol{k})$ and $\hat{d}_y(\boldsymbol{k})$ around the K(1/3, 1/3, L) point (see Supplementary Materials S5 for $\hat{d}_x(\boldsymbol{k})$ and $\hat{d}_y(\boldsymbol{k})$ in the entire Brillouin zone). The top and bottom panels are corresponding to *L*=2 and 1, respectively. Here $\hat{d}_x(\boldsymbol{k})$ and $\hat{d}_y(\boldsymbol{k})$ are obtained by measuring $J(\boldsymbol{k} + \boldsymbol{G})$ at $\boldsymbol{G}$ =(0,0,2) and (0,-1,2) r.l.u. (see Methods). As comparison, we also show $\hat{d}_x(\boldsymbol{k})$ near the Γ( 0, 0, L) points in Fig. 2**d**. Near these high symmetry point, the relation between the $\hat{\boldsymbol{d}}$-vector and experimental azimuthal angle, $\varphi$ (Fig. 2**c**), is $\hat{d}_x(\boldsymbol{k}) = \cos(n\varphi)$ and $\hat{d}_y(\boldsymbol{k}) = \sin(n\varphi)$, where *n* is the winding number of the $\hat{\boldsymbol{d}}$-vector. Theoretically, the $\hat{\boldsymbol{d}}$-vector around the K point lays on the great circle and normal to the surface of Bloch sphere (right inset of Fig. 2**a**), resulting in non-trivial QGT and winding number *n*=1. This prediction is remarkably consistent with the INS measurement shown in Fig. 2**b** and **c**. In contrast, the Γ point features geometrically trivial $\hat{\boldsymbol{d}}$-vectors that is pinned to opposite ends of the Bloch sphere. The corresponding $\hat{d}_x(\boldsymbol{k}) = \pm 1$ is fully consistent with experimental data shown in Fig. 2**d**.

The complete measurement of $\hat{d}_x(\boldsymbol{k})$ and $\hat{d}_y(\boldsymbol{k})$ allows us to construct quantum metric. For a gapless Dirac magnon, the QGT take the simple form with: $g_n^{xx}(\boldsymbol{k}) = [\hat{d}_x(\boldsymbol{k})/|\boldsymbol{q}|]^2$, $g_n^{yy}(\boldsymbol{k}) = [\hat{d}_y(\boldsymbol{k})/|\boldsymbol{q}|]^2$, and $g_n^{xy}(\boldsymbol{k}) = \hat{d}_x(\boldsymbol{k})\hat{d}_y(\boldsymbol{k})/|\boldsymbol{q}|^2$, where $\boldsymbol{q} = \boldsymbol{k} - \boldsymbol{K}$. The $g_n^{ij}(\boldsymbol{k})$ can also be

derived without the knowledge of Dirac-magnon by applying the generic derivative operations described by Eqs. (3) and (4). Figure 3**a**-**c** show the experimentally determined $g_+^{xx}(\boldsymbol{k})$, $g_+^{yy}(\boldsymbol{k})$, and $g_+^{xy}(\boldsymbol{k})$, respectively. The main panel show constructed $g_+^{ij}(\boldsymbol{k})$ based on the gapless Dirac magnon, whereas the insets show experimental $g_+^{ij}(\boldsymbol{k})$ obtained from the generic derivative of $\hat{d}_{\boldsymbol{x}}(\boldsymbol{k})$ and $\hat{d}_{\boldsymbol{y}}(\boldsymbol{k})$. In comparison, Figs. 3**d**-**f** show the theoretically calculated $g_+^{ij}(\boldsymbol{q})$ derived from a $J_1$-$J_2$-$J_3$ model that approximately describes the Dirac magnon in Gd.[33] The overall agreement between different experimental approach and numerical simulation firmly establishes our experimental protocol to measure QGT.

## Discussion

While our case study has focused on gapless Dirac magnon with zero Berry curvature, our method works remarkably well to massive Dirac magnon systems (see Supplementary Materials S6 and S7). Significantly, this model-independent approach does not require knowledge of microscopic magnetic interactions and massive data to extract the effective magnetic Hamiltonian. This generic approach can hence be applied to various quantum magnets including the recently identified Dirac, Weyl, and Chern magnon materials,[36-40] such as $CrI_3$ and $CoTiO_3$. More interestingly, our results bring a new avenue for accessing quantum information encoded in quasiparticle wavefunctions. Fundamentally, the quantum metric of one-magnon state is directly proportional to the quantum Fisher information, $F_n$, with $F_n = 4g_n$ at zero temperature. Our measurements thus provide direct experimental access to a quantity that lies at the heart of quantum metrology and quantum information science. This connection elevates quantum geometry from a mathematical characterization of Bloch wavefunctions to an experimentally accessible information-theoretic resource, linking the sensitivity of quantum states, the topology of energy bands, and the dynamics of collective excitations within a unified framework. Looking forward, extending these measurements to interacting, non-equilibrium, and strongly correlated magnetic systems may reveal new connections between quantum geometry, many-body entanglement, and emergent quantum phases.

## Method:

### 1. Measurement of the $\hat{\boldsymbol{d}}$-vector:

Following Eq. (7-1) of the main text, we can obtain:

$$\begin{pmatrix} \hat{d}_x(\boldsymbol{k}) \\ \hat{d}_y(\boldsymbol{k}) \end{pmatrix} = \frac{1}{\sin(\varphi_{G_1}-\varphi_{G_2})} \begin{pmatrix} -\sin[\varphi_{\boldsymbol{G}_2}] & \sin[\varphi_{\boldsymbol{G}_1}] \\ -\cos[\varphi_{\boldsymbol{G}_2}] & \cos[\varphi_{\boldsymbol{G}_1}] \end{pmatrix} \begin{pmatrix} J(\boldsymbol{k}+\boldsymbol{G}_1) \\ J(\boldsymbol{k}+\boldsymbol{G}_2) \end{pmatrix}, \text{(M.1)}$$

directly from the measurement. With $\hat{d}_x(\boldsymbol{k})$ and $\hat{d}_y(\boldsymbol{k})$ fixed, then $\hat{d}_z(\boldsymbol{k})$ is either $\sqrt{1-\hat{d}_x^2(\boldsymbol{k})-\hat{d}_y^2(\boldsymbol{k})}$ or $-\sqrt{1-\hat{d}_x^2(\boldsymbol{k})-\hat{d}_y^2(\boldsymbol{k})}$. The sign freedom is not random as $\hat{d}_z(\boldsymbol{k})$ is smooth in the Brillouin zone except the band touching points. Then the sign can only change when $\hat{d}_z(\boldsymbol{k})=0$, such as the Dirac point. Then, we can determine the sign change of $\hat{d}_z(\boldsymbol{k})$ by tracking the nodal line of $|\hat{d}_z(\boldsymbol{k})| = \sqrt{1-\hat{d}_x^2(\boldsymbol{k})-\hat{d}_y^2(\boldsymbol{k})}$. As a result, we only need to determine the sign of $\hat{d}_z(\boldsymbol{k})$ at a single high-symmetry point. The formalism can be straightforwardly generalized to cases where two spins are not identical.

**2. Construction of $J(\boldsymbol{Q})$ and the $\hat{\boldsymbol{d}}$-vector:**

To extract the intensities of two magnon branches, $I_+(\boldsymbol{Q})$ and $I_-(\boldsymbol{Q})$ at L=2, we combined the INS measurements acquired at L=1 and L=2. Owing to the L-dependent neutron-scattering intensities, an individual data set may predominantly resolve only one of the two branches, whereas their combination provides access to both modes over the whole momentum space. At each momentum point, the energies of the two modes were determined from the local maximum of the energy-dependent intensity. The corresponding spectral intensities, $I_+(\boldsymbol{Q})$ and $I_-(\boldsymbol{Q})$, were then obtained by integrating the intensity over the energy window of $\pm$FWHM (~1.5 meV) about the respective peak energies.

The INS geometric quantity, $J(\boldsymbol{Q})$, was then constructed directly using $I_+(\boldsymbol{Q})$ and $I_-(\boldsymbol{Q})$ based on Eq. (7-1). The resulting $J(\boldsymbol{Q})$ map over multiple Brillouin zone is shown in Supplementary Materials S5. The pseudospin components $\hat{d}_x(\boldsymbol{k})$and $\hat{d}_y(\boldsymbol{k})$ were subsequently reconstructed from $J(\boldsymbol{Q})$ using Eq. (M.1). Here, $\boldsymbol{k}$ denotes crystal momentum within the first Brillouin zone, whereas $\boldsymbol{Q} = \boldsymbol{k} + \boldsymbol{G}$ denotes the momentum transfer measured around the reciprocal-lattice vector **G**. The x-component is directly obtained from $J(\boldsymbol{k} + \boldsymbol{G})$ with $\boldsymbol{G}$ =(0,0,2) r.l.u.:

$$\hat{d}_x(\boldsymbol{k}) = J(\boldsymbol{k}+\boldsymbol{G})$$

The y-component is obtained by combining $J(\boldsymbol{k} + \boldsymbol{G})$ and $J(\boldsymbol{k} + \boldsymbol{G}')$ with $\boldsymbol{G}'$=(0,-1,2) r.l.u.:

$$\hat{d}_y(\boldsymbol{k}) = \frac{1}{\sqrt{3}}J(\boldsymbol{k} + \boldsymbol{G}) + \frac{2}{\sqrt{3}}J(\boldsymbol{k} + \boldsymbol{G}')$$

The resulting $\hat{d}_x(\boldsymbol{k})$and $\hat{d}_y(\boldsymbol{k})$ map over the Brillouin zone is shown in Supplementary Materials S5.

**3. Construction of quantum metric $g^{ij}$:**

The quantum metric components $g^{ij}$ were obtained from the momentum derivatives of the pseudospin map $\hat{d}_x(\boldsymbol{k})$ and $\hat{d}_y(\boldsymbol{k})$ obtained above following Eq. (3). The experimental $\hat{d}_x(\boldsymbol{k})$ and $\hat{d}_y(\boldsymbol{k})$ maps were first smoothed using a Gaussian filter with $\sigma = 0.8$ pixels to suppress noise during numerical differentiation. The derivatives were then evaluated along the (1,1) and (1,-1) momentum directions using a centered finite-difference scheme. The resulting derivatives were used to obtain $g^{ij}$ shown in insets of Fig. 3.

**Dada availability**: Inelastic neutron scattering data is available at https://doi.org/10.13139/ORNLNCCS/1783161.

**Competing interests**: The authors declare no competing interests.

**Acknowledgements:** We thank Allen Scheie, Garrett Granroth, Tiantian Zhang, Ruixing Zhang, Chiranjit Mondal, Tom Berlijn, Liangbo Liang, and Jiaqiang Yan for stimulating discussions. This research was supported by the U.S. Department of Energy (DOE), Office of Science, Basic Energy Sciences, Materials Sciences and Engineering Division (theory and calculation). This research is also sponsored by the Laboratory Directed Research and Development Program of Oak Ridge National Laboratory, managed by UT-Battelle, LLC, for the US Department of Energy (data analysis).

**Note:** After submitting our manuscript in September 2026, we became aware of a related theoretical study proposing a symmetry-based approach to probe the quantum geometry of magnetic excitations using polarization-resolved Raman spectroscopy, with an application to topological magnons in a honeycomb ferromagnet[41].

**References:**

1. King-Simth, R. D. and Vanderbilt, D. Theory of polarization of crystalline solids. *Phys. Rev. B* **47**, 1651 (1993).
2. Marzari, N. and Vanderbilt, D. Maximally localized generalized Wannier functions for composite energy bands. *Phys. Rev. B* **56**, 12847 (1997).
3. Törmä, P. Essay: where can quantum geometry lead us? *Phys. Rev. Lett.* **131**, 240001 (2023).
4. Liu, T. et al., Quantum geometry in Condensed matter. *National Science Review* **12**, nwae334 (2025).
5. Yu, J. et al., Quantum geometry in quantum materials. *npj Quantum Materials* **10**:101 (2025).
6. Gao, A., Nagaosa, N., Ni, N., and Xu, S.-Y. Quantum geometric phenomena in condensed matter systems. *Rev. Mod. Phys.* doi.org/10.1103/f78t-ky69.
7. Shapere, A. and Wilczek, F. Geometric Phases in Physics. *World Scientific ISBN 9971-50-599-1* (1989).
8. Xiao, D., Chang, M.-C., and Niu, Q. Berry phase effects on electronic properties. *Rev. Mod. Phys.* **82**, 1959 (2010).
9. Haldane, F. D. Model for a quantum Hall effect without Landau levels: condensed matter realization of the "parity anomaly". *Phys. Rev. Lett.* **61**, 2015 (1988).
10. Chang, C.-Z. et al., Experimental observation of the quantum anomalous Hall effect in a magnetic topological insulator. *Science* **340**, 167 (2013).
11. Read, N. and Green, D. Pired states of fermions in two dimensions with breaking of parity and time-reversal symmetries and the fractional quantum Hall effect. *Phs. Rev. B* **61**, 10267 (2000).
12. Kitaev, A. Unpaired Majorana fermions in quantum wires. *Phys.-Usp.* **44**, 131 (2001).
13. Fu, L. and Kane, C. L. Superconducting proximity effect and Majorana Fermions at the surface of a topological insulator. *Phys. Rev. Lett.* **100**, 096407 (2008).
14. Peotta, S. and Törmä, P. Superfluidity in topologically non-trivial flat bands. *Nat. Commun*. **6**:8944 (2015).
15. Tian, H. et al., Evidence for Dirac flat band superconductivity enabled by quantum geometry. *Nature* **614**, 440 (2023).
16. Han, Z. et al., "quantum geometric nesting" and solvable model flat-band systems. *Phys. Rev. X* **14**, 041004 (2024).
17. Gao, Y. and Xiao, D. Nonreciprocal directional dichroism induced by the quantum metric dipole. *Phys. Rev. Lett.* **122**, 227402 (2019).

18. Rhim, J.-W., Kim, K. and Yang, B.-J., Quantum distance and anomalous Landau levels of flat bands. *Nature* **584**, 59 (2020).
19. Hwang, Y., Rhim, J.-W., and Yang, B.-J., Geometric characterization of anomalous Landau levels of isolated flat bands. *Nat. Commun.* **12**:6433 (2021).
20. Wang, N. et al., Quantum-metric-induced nonlinear transport in a topological antiferromagnet. *Nature* **621**, 487 (2023).
21. Gao et al., Quantum metric nonlinear Hall effect in a topological antiferromagnet heterostructure. *Science* **381**, 181-186 (2023).
22. Gao, Y., Yang, S., and Niu, Q. Geometric effects in orbital magnetic susceptibility. *Phys. Rev. B* **91**, 214405 (2015).
23. Piéchon, F. et al., Geometric orbital susceptibility: quantum metric without Berry curvature. *Phys. Rev. B* **94**, 134423 (2016).
24. Lapa, M. F. and Hughes, T. L. Semiclassical wave packet dynamics in nonuniform electric fields. *Phys. Rev. B* **99**, 121111(R) (2019).
25. Onishi, Y. and Fu, L. Emergent curved space and gravitational lensing in quantum materials. *Phys. Rev. B* **113**, 024401 (2026).
26. Shinada, K. and Nagaosa, N. Rienannian geometric classification and emergent phenomena of magnetic textures. arXiv: 2603.05955.
27. Tan, X. et al., Experimental measurement of the quantum metric tensor and related topological phase transition with a superconducting qubit. *Phys. Rev. Lett.* **122**, 210401 (2019).
28. Yu. M, et al., Experimental measurement of the quantum geometric tensor using coupled qubits in diamond. *National Science Review* **7**, 254-260 (2020).
29. Gianfrate, A. et al., Measurement of the quantum geometric tensor and of the anomalous Hall Drift. *Nature* **578**, 281 (2020).
30. Kim, S. et al., Direct measurement of the quantum metric tensor in solids. *Science* **388**, 1050-1054 (2025).
31. Bałut, D., Bradlyn, B., and Abbamonte, P. Quantum entanglement and quantum geometry measured with inelastic x-ray scattering. *Phys. Rev. B* **11**, 125161 (2025).
32. Wu, C., Oka, T., Murakami, S. and Zhang, T. Direct probing the quantum geometric tensor for bosonic collective excitations. arXiv. 2601.13963v3.
33. Scheie, A. et al., Dirac magnons, nodal lines, and nodal plane in elemental gadolinium. *Phys. Rev. Lett*. **128**, 097201 (2022).
34. Scheie, A. et al., Spin-exchange Hamiltonian and topological degeneracies in elemental gadolinium. *Phys. Rev. B* **105**, 104402 (2022).
35. Lindgård, P. A., Harmon, B. N., Freeman, A. J. Theoretical magnon dispersion curves for Gd. *Phys. Rev. Lett*. **35**, 383 (1975).
36. Chen, L. et al., Topological spin excitations in honeycomb ferromagnet $CrI_3$. *Phys. Rev. X* **8**, 041028 (2018).
37. Yuan, B. et al., Dirac magnons in a honeycomb lattice quantum XY magnet CoTiO3. *Phys. Rev. X* **10**, 011062 (2020).

38. Yao, W. et al., Topological spin excitations in a three-dimensional antiferromagnet. *Nat. Phys*. **14**, 1011-1015 (2018).
39. Do, S.-W. et al., Damped Dirac magnon in the metallic kagome antiferromagnet FeSn. *Phys. Rev. B* **105**, L80403 (2022).
40. Do, S.-W. et al., Gaps in topological magnon spectra: intrinsic versus extrinsic effects. *Phys. Rev. B* **106**, L060408 (2022).
41. Yamazaki, Y., Kim, D. Y., Morimoto, T. and Yang, B.-J. Probing the Dirac quantum geometry in quantum magnets from polarized Raman spectroscopy. arXiv: 2611.16491.

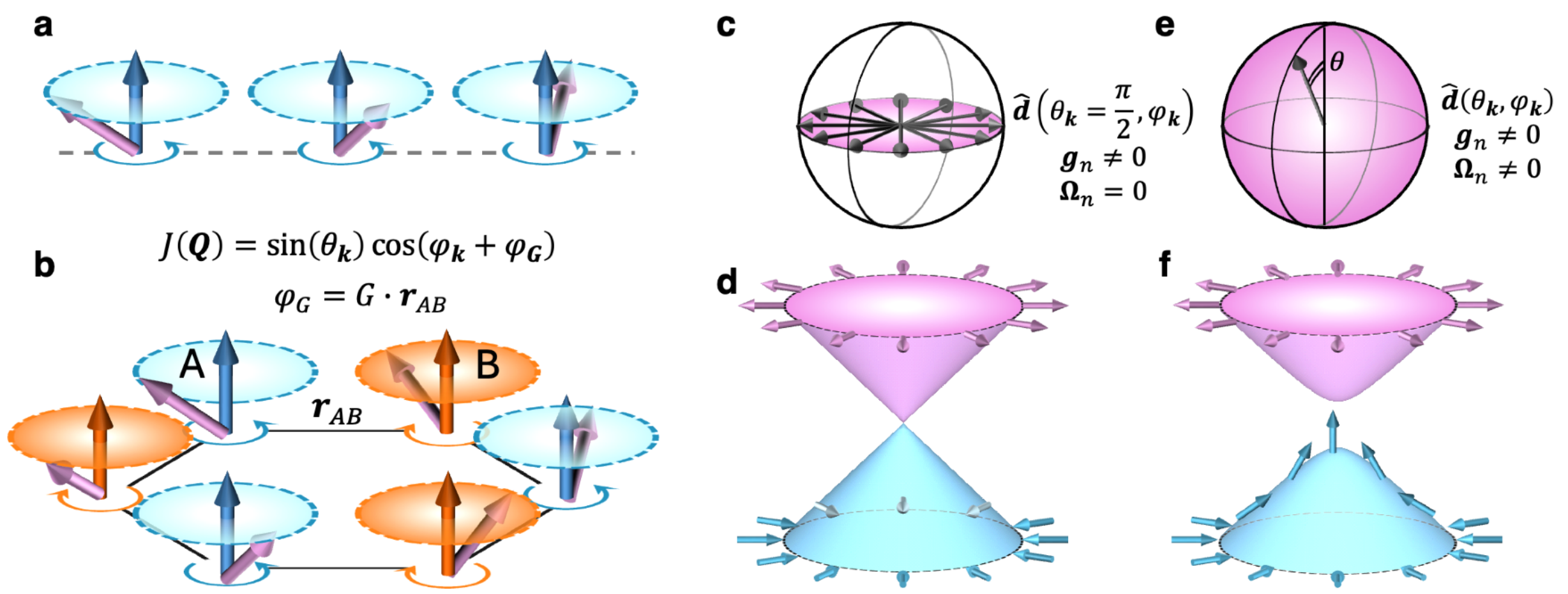

**Figure 1: Magnon quantum geometry. a,** schematic of collective spin procession in a 1D FM chain with one spin in a unit cell. **b,** schematic of collective spin procession in a 2D FM honeycomb lattice with two spins in a unit cell. The magnon wavefunctions are linear combinations of spin processions in A and B sublattice, resulting in two-component complex spinor wavefunctions, $u_{\pm}(\boldsymbol{k})$. The normalized and energy-integrated INS intensity, $J(\boldsymbol{Q})$, only depends on geometric parameters (or the $\hat{\boldsymbol{d}}$-vector) of magnon wavefunction. The spin-sublattice phase vector, $v(\boldsymbol{Q}) = (1, e^{-iG\cdot\boldsymbol{r}_{AB}})/\sqrt{2}$, describes the scattering phase difference between A- and B-sublattices, where $\boldsymbol{r}_{AB} = \boldsymbol{r}_B - \boldsymbol{r}_A$ is the real space translation vector from A-sublattice to B-sublattice (see Supplementary Materials S3). **c**, the $\hat{\boldsymbol{d}}$-vector of the gapless Dirac magnon lays on the great circle of Bloch sphere with $\hat{d}_z(\boldsymbol{k}) = 0$. **d**, the dispersion of gapless Dirac magnon. The red and cyan arrows show the $\hat{\boldsymbol{d}}$-vector of upper and lower branch of magnon bands, respectively. Dirac magnon has non-trivial quantum metric but zero Berry curvature as it doesn't enclose a solid angle of the Bloch sphere. **e**, the $\hat{\boldsymbol{d}}$-vector of magnon Chern band runs over the entire Bloch sphere. **f,** depicts the $\hat{\boldsymbol{d}}$-vector on the upper and lower branch of Chern band.

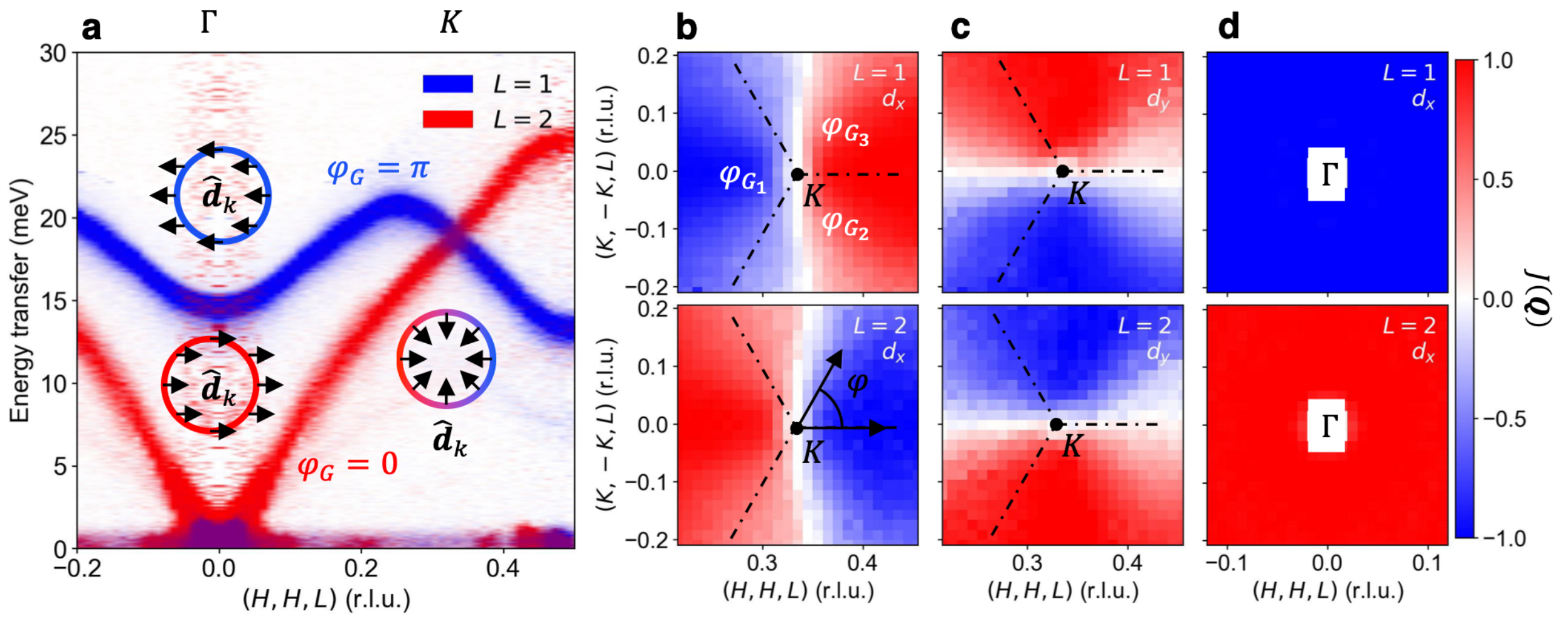

**Figure 2: INS measurement of the geometric $\widehat{\boldsymbol{d}}(\boldsymbol{k})$-vector. a**, $J_{n=\pm}(\boldsymbol{Q},\omega)$ of FM Gd along the (H, H, 0) direction. The red and blue intensity were collected at $L$=2 ± 0.02 and 2 ± 0.02 r.l.u., respectively. Data have been integrated along the (K, -K, 0) direction about K=0±0.02 r.l.u. The left insets and right inset show the $\widehat{\boldsymbol{d}}$-vector near the Γ and K point, respectively. The circles represent magnon eigenvalue contour at a constant energy. **b-c**, experimentally determined $\hat{d}_{\boldsymbol{x}}(\boldsymbol{k})$ and $\hat{d}_{\boldsymbol{y}}(\boldsymbol{k})$ near the K point. Since the K point is located at the Brillouin zone boundary, $J(\boldsymbol{Q})$ measures $\widehat{\boldsymbol{d}}(\boldsymbol{k})$-vector in three Brillouin zones (BZs) with $\varphi_{\boldsymbol{G_1}} = 0$, $\varphi_{\boldsymbol{G_2}} = \frac{2\pi}{3}$, and $\varphi_{\boldsymbol{G_3}} = \frac{4\pi}{3}$, corresponding to $\hat{d}_{\boldsymbol{x}}(\boldsymbol{k})$, $-\frac{1}{2}\hat{d}_{\boldsymbol{x}}(\boldsymbol{k}) - \frac{\sqrt{3}}{2}\hat{d}_{\boldsymbol{y}}(\boldsymbol{k})$, and $-\frac{1}{2}\hat{d}_{\boldsymbol{x}}(\boldsymbol{k}) + \frac{\sqrt{3}}{2}\hat{d}_{\boldsymbol{y}}(\boldsymbol{k})$, respectively. The experimentally determined $\widehat{\boldsymbol{d}}(\boldsymbol{k})$-vector follows $\hat{d}_{\boldsymbol{x}}(\boldsymbol{k}) = \pm\cos(\varphi)$ and $\hat{d}_{\boldsymbol{y}}(\boldsymbol{k}) = \pm\sin(\varphi)$ at $L$=2 (+) and $L$=1 (-), consistent with theoretical prediction. Here $\varphi$ is the experimental azimuth-angle. **d**, shows $\hat{d}_{\boldsymbol{x}}(\boldsymbol{k}) = J(\boldsymbol{k} + \boldsymbol{G_1})$ near the Γ point with $\hat{d}_{\boldsymbol{x}}(\boldsymbol{k})$=1 at $L$=2 (bottom) and -1 at $L$=1 (top), consistent with single-valued $\widehat{\boldsymbol{d}}(\boldsymbol{k})$-vector and trivial quantum geometry. We note that due to the strong magnetic Bragg peak intensity, $J(\boldsymbol{Q})$ close to Γ point is masked.

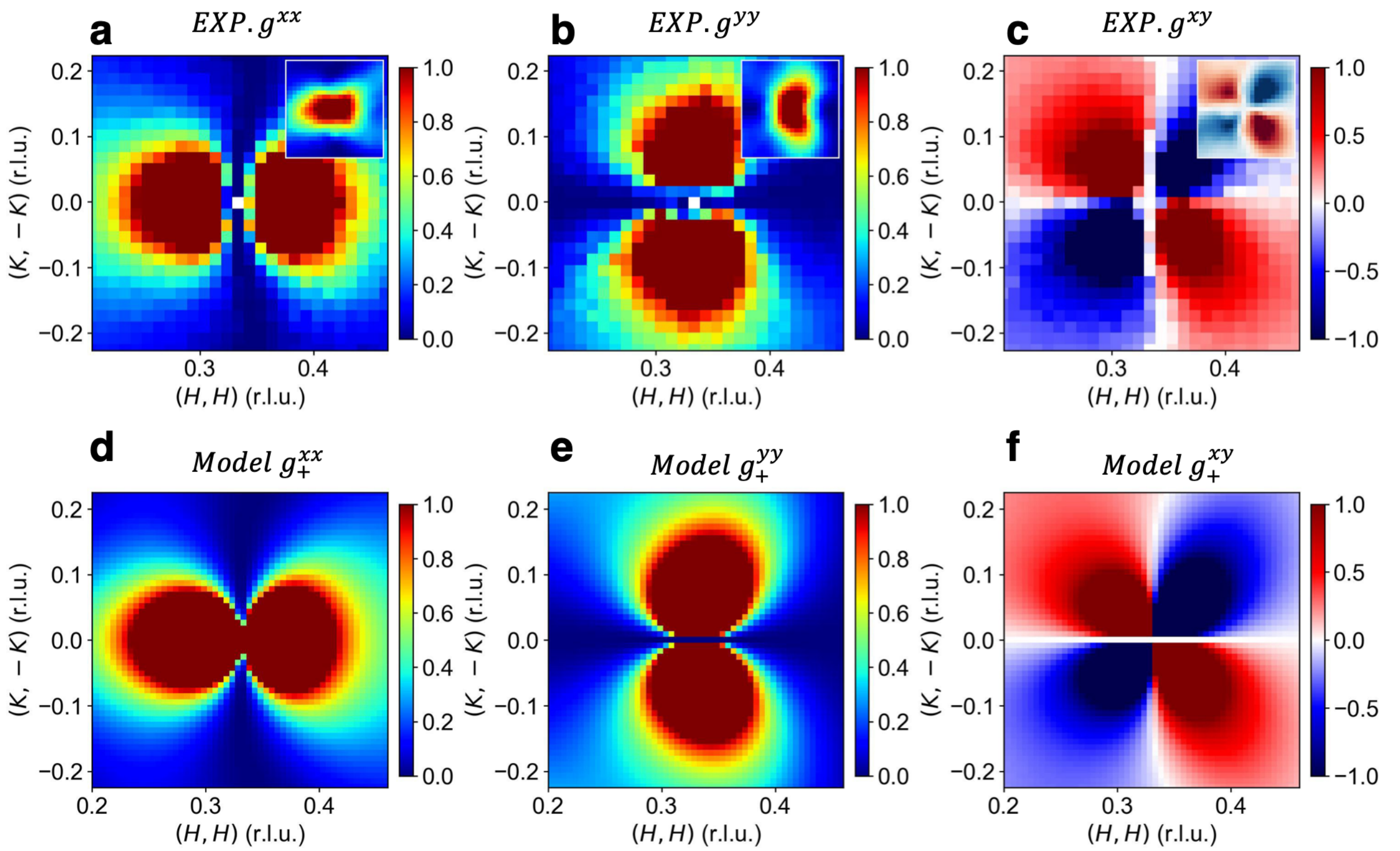

**Figure 3: Magnonic quantum geometry of Gd. a-c**, experimentally determined $g_+^{xx}(\boldsymbol{k}) = [\hat{d}_{\boldsymbol{x}}(\boldsymbol{k})/\boldsymbol{q}]^2$, $g_+^{yy}(\boldsymbol{q}) = [\hat{d}_{\boldsymbol{y}}(\boldsymbol{k})/\boldsymbol{q}]^2$, and $g_+^{xy}(\boldsymbol{k}) = \hat{d}_{\boldsymbol{x}}(\boldsymbol{k})\hat{d}_{\boldsymbol{y}}(\boldsymbol{k})/\boldsymbol{q}^{\mathbf{2}}$ of Gd. Insets show $g_+^{ij}(\boldsymbol{k}) = \frac{1}{4}\partial_{k_i}\widehat{\boldsymbol{d}}(\boldsymbol{k}) \cdot \partial_{k_j}\widehat{\boldsymbol{d}}(\boldsymbol{k})$ based on derivative of $\hat{d}_{\boldsymbol{x}}(\boldsymbol{k})$ and $\hat{d}_{\boldsymbol{y}}(\boldsymbol{k})$ shown in Fig. 2. **d-f,** theoretically calculated $g_+^{ij}$ using a $J_1$-$J_2$-$J_3$ model that produces Dirac magnon at the K point.[33] The remarkable agreements, including the azimuthal dependence and absence of y-mirror symmetry of $g_+^{ij}$, firmly establish a model independent measurement of magnonic quantum geometry.